\documentclass[english]{article}
\usepackage[T1]{fontenc}
\usepackage[latin9]{inputenc}
\usepackage{geometry}
\usepackage{amsmath}
\usepackage{amssymb}
\usepackage{stackrel}
\usepackage{esint}

\makeatletter
\def\Bra#1{\langle #1 |}
\def\Ket#1{|#1 \rangle}
\def\Braket#1{\left\langle\, #1 \,\right\rangle}

\usepackage{babel}

\makeatother

\usepackage{babel}
\begin{document}
\title{Open string field theory with stubs}
\author{Martin Schnabl and Georg Stettinger}
\maketitle
\begin{center}
{\emph{CEICO, Institute of Physics of the Czech Academy of Sciences,
}\\
\emph{ Na Slovance 2, 182 00 Prague 8, Czech Republic}}
\par\end{center}
\begin{abstract}
There are various reasons why adding stubs to the vertices of open
string field theory (OSFT) is interesting: Not only the stubs can
tame certain singularities and make the theory more well-behaved,
but also the new theory shares a lot of similarities with closed string
field theory, which helps to improve our understanding of its structure
and possible solutions. In this paper we explore two natural ways
of implementing stubs into the framework of OSFT, resulting in an
$A_{\infty}$-algebra giving rise to infinitely many vertices. We
find two distinct consistent actions, both generated by a field redefinition,
interestingly sharing the same equations of motion. In the last section
we illustrate their relationship and physical meaning by applying
our construction to nearly marginal solutions. 
\end{abstract}
\tableofcontents{}

\section{Introduction and motivation }

It is a well-known fact that the algebraic and geometric structures
of open and closed string field theory are fundamentally different.
The OSFT action (Witten action) consists of a standard kinetic term
and a single cubic interaction and reads \cite{Witten:1985cc} 
\begin{equation}
S\left(\Psi\right)=\frac{1}{2}\omega\left(\Psi,Q\Psi\right)+\frac{1}{3}\omega\left(\Psi,\Psi\ast\Psi\right).\label{eq:Witten action}
\end{equation}
The algebraic ingredients are a nilpotent differential given by the
BRST-operator, the star product and a symplectic form $\omega$, which
all together form a cyclic differential graded algebra. In the following,
we will use the coalgebra notation of \cite{Vosmera:2020jyw} with
$m_{2}\left(\Psi_{1},\Psi_{2}\right):=\left(-\right)^{d\left(\Psi_{1}\right)}\Psi_{1}\ast\Psi_{2}$
and the shifted degree given by $d\left(\Psi\right)=gh\left(\Psi\right)+1$.
Now the defining algebraic properties are 
\begin{align}
Q^{2} & =0,\\
Qm_{2}\left(\Psi_{1},\Psi_{2}\right) & =-m_{2}\left(Q\Psi_{1},\Psi_{2}\right)-\left(-\right)^{d\left(\Psi_{1}\right)}m_{2}\left(\Psi_{1},Q\Psi_{2}\right),\\
m_{2}\left(\Psi_{1},m_{2}\left(\Psi_{2},\Psi_{3}\right)\right) & =-\left(-\right)^{d\left(\Psi_{1}\right)}m_{2}\left(m_{2}\left(\Psi_{1},\Psi_{2}\right),\Psi_{3}\right),\\
\omega\left(\Psi_{1},m_{2}\left(\Psi_{2},\Psi_{3}\right)\right) & =-\left(-\right)^{d\left(\Psi_{1}\right)}\omega\left(m_{2}\left(\Psi_{1},\Psi_{2}\right),\Psi_{3}\right),\\
\omega\left(\Psi_{1},Q\Psi_{2}\right) & =-\left(-\right)^{d\left(\Psi_{1}\right)}\omega\left(Q\Psi_{1},\Psi_{2}\right).
\end{align}

In contrast, the classical closed string field theory (CSFT) action
\cite{Zwiebach:1992ie} contains infinitely many vertices and reads
\begin{equation}
S\left(\Psi\right)=\frac{1}{2}\omega'\left(\Psi,Q\Psi\right)+\sum_{n=3}^{\infty}\frac{\kappa^{n-2}}{n!}\left\{ \Psi^{n}\right\} ,\label{eq:CSFT action}
\end{equation}
where the vertices 
\begin{equation}
\left\{ \Psi^{n}\right\} =\underset{\mathcal{V}{}_{0,n}}{\int}\Omega_{\Psi^{n}}^{\left(0\right)0,n}
\end{equation}
are given by integrating certain differential forms over vertex regions
$\mathcal{V}{}_{0,n}$ in the moduli space of $n$-punctured Riemann
surfaces. At quantum level there are also vertices of higher genus
to consider, i. e. intrinsic loops already present within the elementary
vertices. However, they will not be of interest here for now.

The vertices also can be decomposed into a symplectic form $\omega'$
\footnote{The symplectic form $\omega'$ differs from the $\omega$ contained
in the Witten action by an additional insertion of $c_{0}^{-}.$} and higher products $l_{n}$: 
\begin{equation}
\left\{ \Psi^{n+1}\right\} =\omega'\left(\Psi,l_{n}\left(\Psi,\Psi,...,\Psi\right)\right).
\end{equation}
In contrast to the open string star product, the $l_{n}$ are all
totally symmetric. The objects $Q=:l_{1},$ all higher $l_{n}$ and
$\omega$ give rise to a cyclic $L_{\infty}$-algebra: 
\begin{align}
 & l_{n}l_{1}\left(\Psi_{1},...,\Psi_{n}\right)+l_{n-1}l_{2}\left(\Psi_{1},...,\Psi_{n}\right)+...+l_{1}l_{n}\left(\Psi_{1},...,\Psi_{n}\right)=0,\,\,\,\,\,\,\,\,\,\forall n\\
 & \omega'\left(\Psi_{1},l_{n}\left(\Psi_{2},...,\Psi_{n+1}\right)\right)=-\left(-\right)^{d\left(\Psi_{1}\right)}\omega'\left(l_{n}\left(\Psi_{1},...,\Psi_{n}\right),\Psi_{n+1}\right).\,\,\,\,\,\,\,\,\,\forall n
\end{align}
The multiplication of the multilinear products in the first line is
defined as 
\begin{equation}
l_{k}l_{l}\left(\Psi_{1},...,\Psi_{k+l-1}\right)=\sum_{\sigma}\frac{\left(-1\right)^{\epsilon\left(\sigma\right)}}{l!\left(k-1\right)!}l_{k}\left(l_{l}\left(\Psi_{\sigma\left(1\right)},...,\Psi_{\sigma\left(l\right)}\right),\Psi_{\sigma\left(l+1\right)},...,\Psi_{\sigma\left(k+l-1\right)}\right),
\end{equation}
where the sum runs over all permutations of $k+l-1$ elements and
the factor $\left(-1\right)^{\epsilon\left(\sigma\right)}$ is just
the sign picked up when permuting the $\Psi$s.

The different algebraic structures have a geometric origin: For a
unitary theory it is necessary that the Feynman string diagrams computed
from the action cover the full moduli space of $n$-punctured Riemann
surfaces exactly once. For the open string where one needs to consider
surfaces with boundaries it has been shown by Zwiebach \cite{Zwiebach:1990az}
that this is indeed the case for the action (\ref{eq:Witten action})
with only one cubic interaction. In contrast, for the closed string,
i. e. surfaces without boundaries, such a simple cubic representation
of the action seems not to be possible, see \cite{Sonoda:1989sj}.

The simplicity of the OSFT action allowed for the discovery of analytic
classical solutions, most importantly the open string tachyon vacuum
\cite{Schnabl:2005gv}. In classical CSFT, no such analytic solutions
have been found yet \cite{Erler:2019loq,Erler:2022agw}. To make some
effort into this direction, we want to propose the following strategy:
If CSFT cannot be simplified easily, maybe we can make OSFT ``more
complicated'' in the sense that its algebraic and geometric structure
resembles that of CSFT? There are two motivations for that: First,
it could help to make the very abstract language of CSFT more intuitive
and tractable. Second, if we can translate the known analytic solutions
to the new, deformed OSFT, it could give an idea how CSFT solutions
might look like. In total there are three steps to do: 
\begin{enumerate}
\item Find a consistent deformation of the Witten OSFT such that its mathematical
structure resembles CSFT 
\item Find analytic solutions of this deformed theory 
\item Make an educated guess for analytic solutions of CSFT 
\end{enumerate}
The first two tasks will be worked out in this paper while the third
one is left for the future.

The deformation we will consider has many additional interesting aspects:
First, OSFT exhibits some issues with singularities or ill-defined
quantities: For example, there exist ``identity-based'' solutions
\cite{Takahashi:2002ez} (for a recent discussion see \cite{Erler:2019vhl})
which solve the equations of motion but do not have a well defined
action. There are hints that these solutions could be better behaved
in the deformed theory. Moreover, the deformed products $M_{2},M_{3},\dots$
defined in the next section could allow for a natural definition of
a Banach type $A_{\infty}$-algebra over the space of string fields.

The paper is organized as follows: In the second section the deformation
using stubs \cite{Zwiebach:1997fe} is discussed in detail, resulting
in an explicit definition of the higher products of the $A_{\infty}$-algebra.
The necessary mathematical ingredients, namely homological perturbation
theory and homotopy transfer, are introduced as well. As a side result,
we give a consistent method of applying homotopy transfer to general
homotopy equivalences which are not deformation retracts. Section
three deals with analytical solutions and the action of the stubbed
theory. Surprisingly, we find two consistent actions with the same
equations of motion, both generated by a field redefinition. The two
field redefinitions, one coming from homological perturbation theory
and one motivated by closed string field theory, are derived explicitly.
The last section is devoted to the physical interpretation of our
results. We apply the whole construction to a suitable class of solutions
and compute the action up to first non-trivial order in the field
expansion.

\section{Deforming the Witten action using stubs}

We have seen that the algebraic structure of CSFT is that of an $L_{\infty}$-algebra,
so naively we should try to find an $L_{\infty}$-based deformation
of OSFT. However, this would imply a commutative product of open strings,
which would hence be fundamentally different from the Witten product.
We still want the deformation be equivalent to Witten theory, so the
best we can do is to aim for an $A_{\infty}$-algebra with products
$M_{n}$ obeying 
\begin{equation}
M_{n}M_{1}\left(\Psi_{1},\ldots,\Psi_{n}\right)+M_{n-1}M_{2}\left(\Psi_{1},\ldots,\Psi_{n}\right)+\ldots+M_{1}M_{n}\left(\Psi_{1},\ldots,\Psi_{n}\right)=0\,\,\,\,\,\,\,\,\,\forall n
\end{equation}
and the multiplication given by 
\begin{align}
M_{k}M_{l}\left(\Psi_{1},\ldots,\Psi_{k+l-1}\right)=M_{k}\left(M_{l}\left(\Psi_{1},\ldots,\Psi_{l}\right),\Psi_{l+1},\ldots,\Psi_{k+l-1}\right)+\nonumber \\
+\left(-\right)^{d\left(\Psi_{1}\right)}M_{k}\left(\Psi_{1},M_{l}\left(\Psi_{2},\ldots,\Psi_{l+1}\right),\Psi_{l+2},\ldots,\Psi_{k+l-1}\right)+ & \cdots+\nonumber \\
+\left(-\right)^{d\left(\Psi_{1}\right)+\ldots+d\left(\Psi_{k-1}\right)}M_{k}\left(\Psi_{1},\ldots,\Psi_{k-1},M_{l}\left(\Psi_{k},\ldots,\Psi_{k+l-1}\right)\right).
\end{align}
From now on we will use the tensor coalgebra notation of \cite{Vosmera:2020jyw}
where the $A_{\infty}$-relations take the simple form 
\begin{equation}
\sum_{k=1}^{n}\mathbf{M_{k}M_{n+1-k}}=0.
\end{equation}
Geometrically, to mimic the situation of CSFT, we would like to have
a three-vertex that does \emph{not }give rise to a full moduli space
cover, such that higher vertices are necessary. One modification that
achieves both of these requirements is to attach \emph{stubs} on the
three vertex, which means, small pieces of propagating strings on
each input. In this way, there will appear some regions of moduli
space that are not covered by Feynman diagrams and hence have to be
taken over by elementary vertices. The higher products which will
give rise to those vertices will indeed form an $A_{\infty}$-algebra.

The Hamiltonian of our CFT which generates time evolution is $L_{0}$,
so the operator which inserts a strip of length $\lambda$ into a
string diagram is $e^{-\lambda L_{0}}$. The natural modification
of the star product which attaches stubs symmetrically on all three
inputs is then given by \cite{Takezaki:2019jkn}: 
\begin{equation}
m_{2}\left(\cdot,\cdot\right)\rightarrow M_{2}\left(\cdot,\cdot\right)=e^{-\lambda L_{0}}m_{2}\left(e^{-\lambda L_{0}}\cdot,e^{-\lambda L_{0}}\cdot\right).\label{eq:stubbed product}
\end{equation}
$M_{2}$ is cyclic with respect to the simplectic form $\omega$ and
$Q$-invariant: 
\begin{equation}
\omega\left(\Psi_{1},M_{2}\left(\Psi_{2},\Psi_{3}\right)\right)=-\left(-\right)^{d\left(\Psi_{1}\right)}\omega\left(M_{2}\left(\Psi_{1},\Psi_{2}\right),\Psi_{3}\right),
\end{equation}
\begin{equation}
QM_{2}\left(\Psi_{1},\Psi_{2}\right)=-M_{2}\left(Q\Psi_{1},\Psi_{2}\right)-\left(-\right)^{d\left(\Psi_{1}\right)}M_{2}\left(\Psi_{1},Q\Psi_{2}\right).
\end{equation}
Those relations follow from $\left[Q,L_{0}\right]=0$ and the fact
that $L_{0}$ is BPZ-even. It is easy to see that $M_{2}$ is not
associative as expected for an $A_{\infty}$-algebra: 
\begin{align}
e^{-\lambda L_{0}}m_{2}\left(e^{-\lambda L_{0}}\Psi_{1},e^{-2\lambda L_{0}}m_{2}\left(e^{-\lambda L_{0}}\Psi_{2},e^{-\lambda L_{0}}\Psi_{3}\right)\right)\nonumber \\
+\,e^{-\lambda L_{0}}m_{2}\left(e^{-2\lambda L_{0}}m_{2}\left(e^{-\lambda L_{0}}\Psi_{1},e^{-\lambda L_{0}}\Psi_{2}\right),e^{-\lambda L_{0}}\Psi_{3}\right) & \neq0.
\end{align}
The task is now to explicitly determine all the higher products and
prove that they satisfy all relevant properties. This will be done
in two ways, first in a formal algebraic way using homological perturbation
theory and second in a more heuristic way using tree diagrams.

\subsection{Elements of homological perturbation theory }

The starting point of homological perturbation theory (HPT) is a \emph{chain
homotopy equivalence}. Let $V$, $W$ be two chain complexes, i. e.
graded vector spaces with a nilpotent differential of degree one denoted
by $d_{V}\left(d_{W}\right)$. Furthermore, we are given two chain
maps $p:\,V\rightarrow W$ and $i:\,W\rightarrow V$ of degree zero
as well as a homotopy map $h:V\rightarrow V$ of degree minus one
obeying the relations 
\begin{align}
pd_{V} & =d_{W}p\\
id_{W} & =d_{V}i\\
ip-1 & =hd_{V}+d_{V}h.
\end{align}
This ensures that $i$ and $p$ as well as the combination $ip$ leave
cohomology classes invariant. 

Let us assume that the differential $d_{V}$ is perturbed by some
$\delta$ of degree one such that $\left(d_{V}+\delta\right)^{2}=0$
again. The \emph{homological perturbation lemma }(HPL) now states
that one can construct a new homotopy equivalence using the perturbed
differential with the other maps given by 
\begin{align}
D_{V} & =d_{V}+\delta\\
D_{W} & =d_{W}+p\delta\left(1-h\delta\right)^{-1}i\\
I & =\left(1-h\delta\right)^{-1}i\\
P & =p\left(1-\delta h\right)^{-1}\\
H & =\left(1-h\delta\right)^{-1}h.
\end{align}
The expression $\left(1-h\delta\right)^{-1}$ requires some explanation:
Usually it should be understood as a geometric series 
\begin{equation}
\left(1-h\delta\right)^{-1}=1+h\delta+h\delta h\delta+h\delta h\delta h\delta+...\,\,\,\,\,\,\,\,\,.
\end{equation}
We demand for the validity of the HPL that this series either converges
or terminates, which will be the case in our examples. The proof of
the lemma including all relevant relations is straightforward but
tedious and can be found in \cite{Crainic,Vosmera:2020jyw}.

An important case in literature is that of a\emph{ deformation retract
}(DR) where $W$ is isomorphic to a subspace of $V$, see e.g. \cite{Erbin:2020eyc}.
The maps $i$ and $p$ are then given by the canonical inclusion and
projection, respectively, hence it follows trivially that $pi=1$.
If additionally the following \emph{annihilation conditions 
\begin{equation}
hh=0,\,\,\,\,\,\,hi=0,\,\,\,\,\,\,ph=0\label{eq:annihilation conditions}
\end{equation}
}are satisfied, it is called a \emph{special deformation retract}
(SDR). It can be shown that any DR can be turned into an SDR by a
redefinition of the maps \cite{Crainic}. If the perturbation lemma
is applied to an DR, the perturbed data will in general not form a
DR again$.$ However, in the case of an SDR, all relations are preserved
after the perturbation, so $PI=1$ and $I$, $P$ and $H$ also fulfill
the annihilation conditions.

\subsection{Transferring algebraic structure: SDR-case}

Our main purpose for using the HPL will be transferring some algebraic
structure from one side of the homotopy equivalence to the other.
Lets consider an SDR with an associative multiplication defined on
$V$. It is a natural question to ask if there exists some ``induced
product'' defined on $W$: For example, for $a,$ $b$ $\in$ $W$
the product $ia\cdot ib$ in $V$ need not be in $W$ anymore. The
only thing that could be done is to project it to $W$, i. e. define
$a\cdot b\mid_{W}=p\left(ia\cdot ib\right)$, but this product will
in general not be associative anymore. What happens is that an associative
algebra on one side of the equivalence becomes an $A_{\infty}$-algebra
after transferring to the other side. A formal way to construct all
the higher products is given by the so-called \emph{tensor trick: }

One defines a new SDR over the tensor coalgebra\footnote{See the appendix for a short summary on tensor coalgebras. }
of $V$ ($W$) with the tensorial versions of the maps given by 
\begin{align}
\mathbf{} & \mathbf{d_{V\left(W\right)}}=\underset{n=1}{\overset{\infty}{\sum}}\underset{k=0}{\overset{n-1}{\sum}}1^{\otimes k}\otimes d_{V\left(W\right)}\otimes1^{\otimes n-k-1}\\
 & \mathbf{i}=\underset{n=1}{\overset{\infty}{\sum}}i^{\otimes n}\\
 & \mathbf{p}=\underset{n=1}{\overset{\infty}{\sum}}p^{\otimes n}\\
 & \mathbf{h}=\underset{n=1}{\overset{\infty}{\sum}}\underset{k=0}{\overset{n-1}{\sum}}1^{\otimes k}\otimes h\otimes\left(ip\right)^{\otimes n-k-1}.
\end{align}
$\mathbf{d_{V\left(W\right)}}$ is the standard tensor coderivation
associated to $d_{V\left(W\right)}$, see (\ref{eq:coderivation}),
$\mathbf{i}$ and $\mathbf{p}$ are cohomomorphisms while $\mathbf{h}$
is defined somewhat asymmetrically which will turn out to be crucial
for the formalism to work. One can check directly that those definitions
indeed give rise to an SDR again. The product on $V$, denoted by
$m_{2}$, can now be treated as a perturbation $\delta$ of the coderivative
$\mathbf{d_{V}}$, 
\begin{equation}
\mathbf{D_{V}}=\mathbf{d_{V}}+\mathbf{m_{2}},
\end{equation}
where $\mathbf{m_{2}}$ is the coderivation associated to $m_{2}$.
$\mathbf{D_{V}}^{2}=0$ will be fulfilled as a consequence of $m_{2}$
being associative and $d_{V}$ obeying the Leibniz rule. According
to the HPL we will be given a new, perturbed complex with the maps
\begin{align}
 & \mathbf{\mathbf{D_{W}}=d_{W}+pm_{2}\left(1-hm_{2}\right)^{-1}i}\\
 & \mathbf{\boldsymbol{I}}=\mathbf{\left(1-hm_{2}\right)^{-1}i}\\
 & \mathbf{\boldsymbol{P}}=\mathbf{\boldsymbol{p}\left(1-m_{2}h\right)^{-1}}\\
 & \mathbf{\boldsymbol{H}}=\mathbf{\boldsymbol{h}\left(1-hm_{2}\right)^{-1}},
\end{align}
where $\mathbf{D_{W}}$ squares to zero and therefore defines an $A_{\infty}$-algebra
on $W$. The higher products can be obtained by expanding $\mathbf{D_{W}}$,
i. e. projecting onto $n$ inputs and one output: 
\begin{equation}
M_{n}=\pi_{1}\mathbf{D_{W}}\pi_{n}.
\end{equation}
For example we get 
\begin{equation}
M_{2}\left(\cdot,\cdot\right)=pm_{2}\left(i\cdot,i\cdot\right)\label{eq:M_2}
\end{equation}
as already anticipated above.

\subsection{Transferring algebraic structure: Non-SDR-case}

We want to apply those concepts now to the problem of attaching stubs
in OSFT. The space $V$ should be given by the space of string fields
$\mathcal{H}_{BCFT}$ with its grading $d\left(\Psi\right)=gh\left(\Psi\right)+1$
and the differential $Q.$ By comparing (\ref{eq:stubbed product})
with (\ref{eq:M_2}) it seems natural to define 
\begin{equation}
i=p=e^{-\lambda L_{0}}.
\end{equation}
However, we see immediately that this definition does not give rise
to a DR: $i$ and $p$ are not an inclusion and projection anymore
and $pi\neq1.$ Still, in principle the HPL holds for arbitrary homotopy
equivalences, so there is a chance to succeed anyway.

Our choice is so far completely symmetric, so lets define $W=V=\mathcal{H}_{BCFT},$
$d_{W}=d_{V}=Q$ and $h_{W}=h_{V}=h$ where we have to solve 
\begin{equation}
e^{-2\lambda L_{0}}-1=hQ+Qh
\end{equation}
The simplest (although not unique) solution for $h$ is 
\begin{equation}
h=\frac{e^{-2\lambda L_{0}}-1}{L_{0}}b_{0}.
\end{equation}
It is important to stress that $h$ is well-behaved also for $L_{0}=0$
and does not have any pole.

One could now try to proceed with the tensor trick as above and eventually
compute the map $\mathbf{D_{W}}$, but this runs into problems: Although
$\mathbf{D_{W}}^{2}=0$ still, as guaranteed by the HPL, $\mathbf{D_{W}}$
is not a coderivation anymore! This means, it does not obey the co-Leibniz
rule (\ref{eq:co leibniz}). As explained in the Appendix of \cite{Konopka:2015tta},
the condition $pi=1$ as well as the annihilation relations are necessary
(and sufficient) for the tensor trick to work. So there has to be
some modification to account for that and it will actually turn out
to be surprisingly simple: We can just take the expression we get
for $\mathbf{D_{W}}$ and while expanding to calculate the higher
products, \emph{pretend }that all SDR-relations (\ref{eq:annihilation conditions})
are satisfied. More precisely, we define an operator $P_{SDR}$ acting
on the space of maps from $T\mathcal{H}\rightarrow T\mathcal{H}$
which projects on maps in which all SDR-relations hold. This means,
every time that $pi$ occurs, it will be replaced by $1$ and every
time $hh,$ $hi$ or $ph$ occurs, the term will be discarded: 
\[
P_{SDR}\left(....hh....\right)=0,\,\,\,\,\,\,\,P_{SDR}\left(....hi....\right)=0,\,\,\,\,\,\,\,\,P_{SDR}\left(....ph....\right)=0,\,\,\,\,\,\,\,\,P_{SDR}\left(....pi....\right)=P_{SDR}\left(....1....\right)
\]
Now our new higher products will be given as 
\begin{equation}
M_{n}=P_{SDR}\pi_{1}\mathbf{D_{W}}\pi_{n}.
\end{equation}
Their associated coderivations can be added together to form a total
map called $\mathbf{M}$, 
\begin{equation}
\mathbf{M}=\underset{n=1}{\overset{\infty}{\sum}}\underset{m=n}{\overset{\infty}{\sum}}\underset{k=0}{\overset{m-n}{\sum}}1^{\otimes k}\otimes M_{n}\otimes1^{\otimes m-k-n}.
\end{equation}
It is a coderivation by construction and moreover it squares to zero
because in the proof of the HPL, where it is shown that $D_{W}^{2}=0$
(Eq. 1.2.16 of \cite{Vosmera:2020jyw}), the SDR-relations are never
used. Since we know that $\mathbf{D_{W}}^{2}=0$ independently of
the SDR-relations, also $\mathbf{M}^{2}=0$ and $\mathbf{M}$ defines
the desired $A_{\infty}$-algebra.

As an example, lets explicitly calculate $M_{3}:$ 
\begin{align}
P_{SDR}\pi_{1}\mathbf{D_{W}}\pi_{3} & =P_{SDR}\pi_{1}\mathbf{pm_{2}h\mathbf{m_{2}}i}\pi_{3}=P_{SDR}\left(pm_{2}\left(1\otimes h+h\otimes ip\right)\left(1\otimes m_{2}+m_{2}\otimes1\right)i^{\otimes3}\right)\nonumber \\
 & =P_{SDR}\left(pm_{2}\left(i\cdot,hm_{2}\left(i\cdot,i\cdot\right)\right)-pm_{2}\left(m_{2}\left(i\cdot,i\cdot\right),hi\cdot\right)+pm_{2}\left(hi\cdot,ipm_{2}\left(i\cdot,i\cdot\right)\right)+pm_{2}\left(hm_{2}\left(i\cdot,i\cdot\right),ipi\cdot\right)\right)\nonumber \\
 & =pm_{2}\left(i\cdot,hm_{2}\left(i\cdot,i\cdot\right)\right)+pm_{2}\left(hm_{2}\left(i\cdot,i\cdot\right),i\cdot\right).
\end{align}
The second and third term in the second line contained an $hi$ and
were deleted whereas in the last term $ipi$ was replaced by $i$.

This whole procedure including the proof of the statement may seem
quite handwavy, however, there exists a more formal and precise way
of arriving at the same result using operad theory \cite{Loday}.
A combinatorial proof using tree diagrams will be given in the next
section.

\subsection{Higher products using tree diagrams}

The proposal is that $M_{n}$ is equal to the sum of all distinct,
rooted, full binary trees with $n$ leaves such that every leaf represents
one input and the root is the output. With every leaf there is one
factor of $i$ associated, with every node the product $m_{2}$, with
every internal line $h$ and with the root $p$. In \cite{Li:2018rnc}
it is argued that this is true for SDRs, since we construct the products
in the same way as for an SDR, we conclude that the proposal also
holds for our non-SDR case.

In the tree language, the $A_{\infty}$-relations can be proven directly:
The commutator of $\mathbf{Q}$ with an $n$-leaved tree gives a sum
of $n-2$ terms, in which one of the $n-2$ internal lines $h$ is
replaced by $1-ip$. The $1$-terms actually cancel away because of
the associativity of $m_{2}$: The propagator associated with unity
connects two nodes $m_{2}$ which leads to an expression of the form
$m_{2}\left(m_{2}\left(A,B\right),C\right)$ where $A,B,C$ are three
subtrees. In the sum there always exists a second tree with another
propagator turned into unity giving rise to the expression $m_{2}\left(A,m_{2}\left(B,C\right)\right)$.
These two trees cancel away such that only the -$ip$-factors remain
in total. Now the other terms occuring in the relation 
\begin{equation}
-\left[\mathbf{Q,M_{n}}\right]=\mathbf{M_{2}M_{n-1}}+\mathbf{M_{3}M_{n-2}}+...+\mathbf{M_{n-1}M_{2}}\label{eq:A infinity again}
\end{equation}
can be interpreted as follows: If we project on one output, $\pi_{1}\mathbf{M_{k}M_{n+1-k}}$
gives a sum of trees where one of the terms in $M_{n+1-k}$ is connected
with its root to one of the $k$ leaves of one of the trees in $M_{k}$
and this is done in all possible combinations. The result is a sum
of trees with in total $n$ leaves, where one of the internal lines
does not contain $h$ but $ip$; the $i$ from the leaf of the left
tree and the $p$ from the root of the right tree. But that is exactly
the same sum of terms we have on the l.h.s., indeed it is easy to
see that each tree occuring on the r.h.s. must also occur on the l.h.s.
and vice versa.

An interesting crosscheck of the $\pi_{1}$-projection of (\ref{eq:A infinity again})
can be done by comparing the total number of trees on both sides:
The number of full binary trees with $n+1$ leaves is given by the
Catalan number 
\begin{equation}
C_{n}=\frac{1}{n+1}\begin{pmatrix}2n\\
n
\end{pmatrix}.
\end{equation}
This means that on the l.h.s. there are $C_{n-1}$ trees in $M_{n}$
and $n-2$ internal lines that can be changed, hence a total of $C_{n-1}\left(n-2\right)$
trees. On the r.h.s. we have $2\cdot C_{1}C_{n-2}+3\cdot C_{2}C_{n-3}+...+\left(n-1\right)C_{n-2}C_{1}$
trees in total, leading to the equation 
\begin{equation}
\overset{n-1}{\underset{k=2}{\sum}}\left(n+1-k\right)C_{n-k}C_{k-1}=\left(n-2\right)C_{n-1}.\label{eq:Catalan relation}
\end{equation}
The Catalan numbers fulfill the following useful recursive relations:
\begin{equation}
C_{n+1}=\frac{2\left(2n+1\right)}{n+2}C_{n},\,\,\,\,\,\,\,\,\,\,C_{n+1}=\overset{n}{\underset{k=0}{\sum}}C_{n-k}C_{k}.
\end{equation}
Using them one can proceed by induction: Assuming equation (\ref{eq:Catalan relation})
is valid for some $n$, then 
\begin{align}
 & \overset{n}{\underset{k=2}{\sum}}\left(n+2-k\right)C_{n+1-k}C_{k-1}\nonumber \\
= & \,\overset{n}{\underset{k=2}{\sum}}\left(n+2-k\right)\frac{2\left(2n-2k+1\right)}{n-k+2}C_{n-k}C_{k-1}\nonumber \\
= & \,4\overset{n-1}{\underset{k=2}{\sum}}\left(n+1-k\right)C_{n-k}C_{k-1}-2\overset{n-1}{\underset{k=2}{\sum}}C_{n-k}C_{k-1}+2C_{0}C_{n-1}\nonumber \\
= & \left(4n-8\right)C_{n-1}-2\overset{n-2}{\underset{k=1}{\sum}}C_{n-1-k}C_{k}+2C_{n-1}\nonumber \\
= & \left(4n-6\right)C_{n-1}-2\left(C_{n}-C_{0}C_{n-1}-C_{n-1}C_{0}\right)\nonumber \\
= & \left(4n-2\right)\frac{n+1}{2\left(2n-1\right)}C_{n}-2C_{n}\nonumber \\
= & \left(n-1\right)C_{n}
\end{align}
as it should be to complete the induction. This shows that the number
of terms in the equation (\ref{eq:A infinity again}) is the same
on both sides.

\subsection{Proof of cyclicity to all orders}

Using the tree language it is possible to prove that all higher products
$M_{n}$ are cyclic with respect to the BPZ-product. We have to show
\begin{equation}
\omega\left(\Psi_{1},M_{n}\left(\Psi_{2},...,\Psi_{n+1}\right)\right)=-\left(-\right)^{d\left(\Psi_{1}\right)}\omega\left(M_{n}\left(\Psi_{1},...,\Psi_{n}\right),\Psi_{n+1}\right),\label{eq:cyclicity}
\end{equation}
hence we start with a sum of trees on the l.h.s. and use the BPZ-properties
of $m_{2}$ and $h$ as well as $p=i^{\dagger}=i$ to rewrite it as
the sum of trees on the r.h.s.. The explicit steps are: 
\begin{enumerate}
\item Take the $p$ from the root of the tree and write it to the left side
of the product where it can be interpreted as $i$, acting on $\Psi_{1}$. 
\item Take the $m_{2}$ from the root of the tree and use cyclicity of $m_{2}$
to apply it on the first two arguments inside of $\omega$. This gives
a sign factor of $-\left(-\right)^{d\left(\Psi_{1}\right)}$. In general
one will be left with two subtrees then with $n+1$ leaves in total. 
\item Take the $h$ from the right subtree and use that it is BPZ-even to
apply it on the left subtree. This gives an additional sign factor
of $\left(-\right)^{d\left(left\,\,subtree\right)}$. 
\item Take the $m_{2}$ from the root of the right subtree and apply it
on the first two arguments inside of $\omega$. It gives a sign factor
of $-\left(-\right)^{d\left(left\,\,subtree\right)+1}$ (the $+1$
comes from the $h$ that was shifted in step 3) which cancels the
sign factor of step 3. Again, one is left with two subtrees. 
\item Repeat steps 3 and 4 until the right subtree only consist of $i$
acting on one input. The total sign factor remains $-\left(-\right)^{d\left(\Psi_{1}\right)}$. 
\item Remove the $i$ acting on $\Psi_{n+1}$ and let it act as a $p$ on
the left subtree. 
\end{enumerate}
Now the left subtree fulfills all requirements to be an element of
$M_{n}$ and since we also have the right sign factor, we have obtained
a term contained in the r.h.s. of (\ref{eq:cyclicity}). The manipulations
are all uniquely invertible so we can conclude that the map between
the trees is one-to-one and all terms we need are constructed exactly
once. As a result, Eq. (\ref{eq:cyclicity}) holds and all higher
products are cyclic. Moreover, as already suggested by the name, Eq.
(\ref{eq:cyclicity}) together with the antisymmetry of $\omega$
implies invariance of the vertices under cyclic permutations.

\subsection{Geometric picture}

We have now shown that the higher products fulfill all the algebraic
requirements but we do not know anything yet about the geometric picture,
if they indeed give rise to a full single cover of the moduli space.
To answer this question, the tree description of the products turns
out to be very useful. Lets consider an arbitrary string tree diagram
using the stubbed three vertex: As long as the external states are
on-shell, the stubs make no difference because the external legs consist
of a semiinfinite strip anyway. On the internal lines instead, the
stubs make a difference because all internal strips with a length
smaller than $2\lambda$ do not appear. We can conclude that the additional
elemantary vertices we need should consist of all tree diagrams with
all internal strips having a length smaller than $2$$\lambda$. The
Siegel-gauge string propagator in the Schwinger parametrization is
given by 
\begin{equation}
\int_{0}^{\infty}dt\,e^{-tL_{0}}b_{0}=\frac{b_{0}}{L_{0}}.
\end{equation}
The integral over $t$ can be thought of an integral over strips of
propagating strings of all different lengths. Following this logic,
the propagators in our new vertices should be given as 
\begin{equation}
\int_{0}^{2\lambda}dt\,e^{-tL_{0}}b_{0}=-\frac{e^{-2\lambda L_{0}}-1}{L_{0}}b_{0}=-h
\end{equation}
and hence be equal to minus the homotopy!\footnote{It is interesting to notice at this point that the simple but not
unique choice of $h$ corresponds to choosing Siegel gauge for the
propagator. The minus sign is just a convention and can be absorbed
in the definition of $h$.} We have constructed the higher products by drawing all binary tree
diagrams with Witten vertices, $h$ as propagators and $e^{-\lambda L_{0}}$
on the leaves and the root. Those are in one-to-one correspondence
with all the Feynman tree diagrams that should make up the new elementary
vertices. This shows that our higher products $M_{n}$ indeed define
a set of vertices which gives rise to a full cover of the moduli space.

\section{Analytic solutions and action(s)}

\subsection{Projection cohomomorphism from the HPL}

Using the definition 
\begin{equation}
\mathbf{m}=\mathbf{Q}+\mathbf{m_{2}}
\end{equation}
the equations of motion of the original Witten theory can be written
in coalgebra language as 
\begin{equation}
\mathbf{m}\frac{1}{1-\Psi}=0,
\end{equation}
i. e. solutions are Maurer-Cartan elements of the $A_{2}$-algebra
defining the theory. In the same spirit, to find solutions of the
deformed theory, we have to solve the equation 
\begin{equation}
\mathbf{M}\frac{1}{1-\Psi}=0.
\end{equation}
In fact, the homological perturbation lemma already gave us an operator
which maps solutions of the Witten theory to solutions of the stubbed
theory. This can be seen as follows: The perturbed projection $\mathbf{P}$
is a chain map and hence obeys 
\begin{equation}
\mathbf{Pm=MP.}\label{eq:chain map}
\end{equation}
Now lets assume $\Psi^{\ast}$ is a solution of the Witten theory,
then 
\begin{equation}
\left(\Psi^{*}\right)'=\pi_{1}\mathbf{P}\frac{1}{1-\Psi^{*}}
\end{equation}
obeys 
\begin{equation}
\mathbf{M}\frac{1}{1-\left(\Psi^{*}\right)'}=\mathbf{M}\frac{1}{1-\pi_{1}\mathbf{P}\frac{1}{1-\Psi^{*}}}=\mathbf{MP}\frac{1}{1-\Psi^{*}}=\mathbf{Pm}\frac{1}{1-\Psi^{*}}=0,
\end{equation}
where Eq. (\ref{eq:cohomo relation}) was used. It remains to determine
the cohomorphism $\mathbf{P}$ for the case where the homotopy equivalence
is not an SDR. Again, as for the higher products above, the simplest
way is to take the expression from the HPL 
\begin{equation}
\mathbf{P}_{HPL}=\mathbf{p}\left(\mathbf{1-m_{2}h}\right)^{-1}
\end{equation}
and apply the operator $P_{SDR}$ on its components to get 
\begin{equation}
P_{n}=P_{SDR}\pi_{1}\mathbf{P}_{HPL}\pi_{n}.
\end{equation}
The resulting maps can then be packaged into a cohomorphism $\mathbf{P}$
again. More explicitly, we get for the first few orders 
\begin{align}
P_{1} & =p\nonumber \\
P_{2} & =pm_{2}\left(\cdot,h\cdot\right)+pm_{2}\left(h\cdot,ip\cdot\right)\nonumber \\
P_{3} & =pm_{2}\left(\cdot,hm_{2}\left(\cdot,h\cdot\right)\right)+pm_{2}\left(\cdot,hm_{2}\left(h\cdot,ip\cdot\right)\right)+pm_{2}\left(h\cdot,hm_{2}\left(ip\cdot,ip\cdot\right)\right)+pm_{2}\left(hm_{2}\left(\cdot,h\cdot\right),ip\cdot\right)\nonumber \\
 & \,\,\,\,\,\,+pm_{2}\left(hm_{2}\left(h\cdot,ip\cdot\right),ip\cdot\right)+pm_{2}\left(h\cdot,ipm_{2}\left(\cdot,h\cdot\right)\right)+pm_{2}\left(h\cdot,ipm_{2}\left(h\cdot,ip\cdot\right)\right)\label{eq:interacting Ps}\\
 & ...\,\,\,\,\,\,\,\,\,\,\,\,\,\,.\nonumber 
\end{align}
One can now check the equation $\pi_{1}\mathbf{Pm}=\pi_{1}\mathbf{MP}$
order by order: 
\begin{align}
\pi_{1}\mathbf{Pm}\pi_{1} & =pQ=Qp=\pi_{1}\mathbf{MP}\pi_{1}\nonumber \\
\pi_{1}\mathbf{Pm}\pi_{2} & =P_{2}\left(Q\cdot,\cdot\right)+P_{2}\left(\cdot,Q\cdot\right)+pm_{2}\nonumber \\
 & =-pm_{2}\left(Q\cdot,h\cdot\right)+pm_{2}\left(hQ\cdot,ip\cdot\right)+pm_{2}\left(\cdot,hQ\cdot\right)+pm_{2}\left(h\cdot,ipQ\cdot\right)+pm_{2}\nonumber \\
 & =-pm_{2}\left(Q\cdot,h\cdot\right)-pm_{2}\left(Qh\cdot,ip\cdot\right)+pm_{2}\left(\left(ip-1\right)\cdot,ip\cdot\right)\nonumber \\
 & \,\,\,\,\,-pm_{2}\left(\cdot,Qh\cdot\right)+pm_{2}\left(\cdot,\left(ip-1\right)\cdot\right)+pm_{2}\left(h\cdot,Qip\cdot\right)+pm_{2}\nonumber \\
 & =Qpm_{2}\left(\cdot,h\cdot\right)+Qpm_{2}\left(h\cdot,ip\cdot\right)+pm_{2}\left(ip\cdot,ip\cdot\right)\nonumber \\
 & =QP_{2}+M_{2}\left(p\cdot,p\cdot\right)=\pi_{1}\mathbf{MP}\pi_{2}.
\end{align}
For order three the calculation is already very tedious but it also
turns out to work. The important point is that in the manipulations
that are necessary, the SDR-relations were never used. This implies,
since we know that (\ref{eq:interacting Ps}) works for SDRs, it also
works in our case and (\ref{eq:interacting Ps}) is a valid definition.
Now we have a cohomorphism by construction that obeys the chain map
relation (\ref{eq:chain map}) such that we can construct analytic
solutions of the deformed theory.\footnote{It is actually a non-trivial counting problem to determine the number
of terms of $P_{n}$. It grows faster than the Catalan numbers because
at order three we have seven terms and at order four already 33.}

\subsection{The action}

The on-shell action is one of the most important observables in OSFT,
for example for the tachyon vacuum its value is equal to minus the
energy of the D-brane which has decayed. Since the stubbed theory
should be physically equivalent to the original Witten theory, we
expect that the values for the on-shell action we get in the two theories
coincide. The Witten action can be written in coalgebra notation \cite{Vosmera:2020jyw}
as 
\begin{equation}
S\left(\Psi\right)=\int_{0}^{1}dt\,\omega\left(\pi_{1}\boldsymbol{\partial_{t}}\frac{1}{1-\Psi\left(t\right)},\pi_{1}\mathbf{m}\frac{1}{1-\Psi\left(t\right)}\right),
\end{equation}
where $\Psi\left(t\right)$ is any smooth interpolation between $\Psi\left(0\right)=0$
and $\Psi\left(1\right)=\Psi$ and $\boldsymbol{\partial_{t}}$ the
coderivation associated to $\partial_{t}$. Similarly, the stubbed
action reads 
\begin{equation}
S'\left(\Psi\right)=\int_{0}^{1}dt\,\omega\left(\pi_{1}\boldsymbol{\partial_{t}}\frac{1}{1-\Psi\left(t\right)},\pi_{1}\mathbf{M}\frac{1}{1-\Psi\left(t\right)}\right).\label{eq:deformed action}
\end{equation}
We would expect now a relation 
\begin{equation}
S'\left(\left(\Psi^{*}\right)'\right)\stackrel{?}{=}S\left(\Psi^{*}\right)
\end{equation}
with $\Psi^{*}$ a MC-element of $\mathbf{m}.$ However, this relation
turns out to be non-obvious: Instead 
\begin{align}
S\left(\Psi\right) & =\int_{0}^{1}dt\,\omega\left(\pi_{1}\boldsymbol{\partial_{t}}\frac{1}{1-\Psi\left(t\right)},\pi_{1}\mathbf{m}\frac{1}{1-\Psi\left(t\right)}\right)=\int_{0}^{1}dt\,\omega\left(\pi_{1}\boldsymbol{\partial_{t}}\mathbf{P^{-1}P}\frac{1}{1-\Psi\left(t\right)},\pi_{1}\mathbf{P^{-1}MP}\frac{1}{1-\Psi\left(t\right)}\right)\nonumber \\
 & =\int_{0}^{1}dt\,\omega\left(\pi_{1}\boldsymbol{\partial_{t}}\mathbf{P^{-1}}\frac{1}{1-\pi_{1}\mathbf{P}\frac{1}{1-\Psi\left(t\right)}},\pi_{1}\mathbf{P^{-1}M}\frac{1}{1-\pi_{1}\mathbf{P}\frac{1}{1-\Psi\left(t\right)}}\right)\nonumber \\
 & =\int_{0}^{1}dt\,\omega\left(\pi_{1}\mathbf{P^{-1}}\boldsymbol{\partial_{t}}\frac{1}{1-\left(\Psi\right)'\left(t\right)},\pi_{1}\mathbf{P^{-1}M}\frac{1}{1-\left(\Psi\right)'\left(t\right)}\right)=:\tilde{S}\left(\Psi'\right),\label{eq:action transformation}
\end{align}
where the last line differs from $S'\left(\Psi'\right)$ by the insertions
of $\mathbf{P^{-1}}$ on both inputs of $\omega$.

The invertibility of $\mathbf{P}$ is actually a delicate question:
In general, a cohomomorphism is invertible iff its linear component,
i. e. $P_{1}=p$ is invertible. Now $p=e^{-\lambda L_{0}}$ inserts
a strip of length $\lambda$, so one would expect the inverse $e^{\lambda L_{0}}$
to remove a strip of length $\lambda$ from the world sheet, which
is not always possible. On the other hand, $e^{\lambda L_{0}}$ makes
sense on a string field expanded in eigenstates of $L_{0}$ as long
as the eigenvalues are finite. From now on, we shall assume that this
is the case and $e^{\lambda L_{0}}$ is well-defined on all string
fields in question.

Explicitly, the inversion of $\mathbf{P}$ works as follows: $\mathbf{PP^{-1}}=\mathbf{P^{-1}P}$
should be equal to the identity cohomomorphism, which is identity
in its linear component and zero in all higher components. One can
now solve the components $P_{n}^{-1}=\pi_{1}\mathbf{P^{-1}}\pi_{n}$
order by order: 
\begin{align}
P_{1}^{-1} & =p^{-1}\nonumber \\
P_{2}^{-1} & =-m_{2}\left(p^{-1}\cdot,hp^{-1}\cdot\right)-m_{2}\left(hp^{-1}\cdot,i\cdot\right)\nonumber \\
P_{3}^{-1} & =-m_{2}\left(hp^{-1}\cdot,hm_{2}\left(i\cdot,i\cdot\right)\right)+m_{2}\left(m_{2}\left(p^{-1}\cdot,hp^{-1}\cdot\right),hp^{-1}\cdot\right)+m_{2}\left(m_{2}\left(hp^{-1}\cdot,i\cdot\right),hp^{-1}\cdot\right)\nonumber \\
 & ...\,\,\,\,\,\,\,\,\,\,\,\,\,.
\end{align}
One can now write out in more detail: 
\begin{align}
\tilde{S}\left(\Psi\right)=\, & \frac{1}{2}\omega\left(p^{-1}\Psi,p^{-1}Q\Psi\right)\nonumber \\
+\, & \frac{1}{3}\omega\left(p^{-1}\Psi,p^{-1}M_{2}\left(\Psi,\Psi\right)\right)+\frac{1}{3}\omega\left(p^{-1}\Psi,P_{2}^{-1}\left(\left(Q\Psi,\Psi\right)+\left(\Psi,Q\Psi\right)\right)\right)+\frac{1}{3}\omega\left(P_{2}^{-1}\left(\Psi,\Psi\right),p^{-1}Q\Psi\right)\nonumber \\
+\, & \frac{1}{4}\omega\left(p^{-1}\Psi,p^{-1}M_{3}\left(\Psi,\Psi,\Psi\right)\right)+\frac{1}{4}\omega\left(p^{-1}\Psi,P_{2}^{-1}\left(\left(M_{2}\left(\Psi,\Psi\right),\Psi\right)+\left(\Psi,M_{2}\left(\Psi,\Psi\right)\right)\right)\right)\nonumber \\
+\, & \frac{1}{4}\omega\left(p^{-1}\Psi,P_{3}^{-1}\left(\left(Q\Psi,\Psi,\Psi\right)+\left(\Psi,Q\Psi,\Psi\right)+\left(\Psi,\Psi,Q\Psi\right)\right)\right)+\frac{1}{4}\omega\left(P_{2}^{-1}\left(\Psi,\Psi\right),p^{-1}M_{2}\left(\Psi,\Psi\right)\right)\nonumber \\
+\, & \frac{1}{4}\omega\left(P_{2}^{-1}\left(\Psi,\Psi\right),P_{2}^{-1}\left(\left(Q\Psi,\Psi\right)+\left(\Psi,Q\Psi\right)\right)\right)+\frac{1}{4}\omega\left(P_{3}^{-1}\left(\Psi,\Psi,\Psi\right),p^{-1}Q\Psi\right)+\mathcal{O}\left(\Psi^{\otimes5}\right).
\end{align}

It seems that the cohomomorphism $\mathbf{P}$ does not define the
field redefinition we were looking for, instead it relates the Witten
theory to a theory defined by $\tilde{S}\left(\Psi\right)$. The equations
of motion derived from $\tilde{S}$ are the same as for $S'$, namely
$\mathbf{M}\frac{1}{1-\Psi}=0$, but it is not at all obvious that
the two actions agree even on-shell.

The reason is that the HPL does not know anything about the symplectic
form $\omega$: To get an invariant action with $S'\left(\Psi'\right)\stackrel{}{=}S\left(\Psi\right)$,
not only $\mathbf{m}$ has to transform accordingly, but also $\omega$
would have to go to 
\begin{equation}
\omega'\left(\Psi_{1},\Psi_{2}\right)=\omega\left(\pi_{1}\mathbf{P^{-1}}\frac{1}{1-\Psi_{1}},\pi_{1}\mathbf{P^{-1}}\frac{1}{1-\Psi_{2}}\right),
\end{equation}
otherwise $\mathbf{P}$ would fail to be cyclic. As it can be seen
from (\ref{eq:action transformation}), $\tilde{S}\left(\Psi\right)$
is just a fancy rewriting of the Witten action and therefore defines
an equivalent theory. But since we would like to keep the original
$\omega$, the field redefinition induced by $\mathbf{P}$ from the
HPL and giving rise to $\tilde{S}\left(\Psi\right)$ is not exactly
what we want. However, since it shares the same equations of motion
as $S'\left(\Psi\right)$, it might provide a new family of gauge-invariant
observables for solutions of $S'\left(\Psi\right)$, parametrized
by $\lambda.$ In the last section we will check this explicitly on
a special class of solutions. For now the next task is to derive the
originally desired field redefinition which relates the actions $S\left(\Psi\right)$
and $S'\left(\Psi\right)$. \footnote{It would also be an interesting direction to examine if there exists
some kind of ``dual'' HPL that directly yields this correct field
redefinition.}

\subsection{Elements from closed string field theory}

In \cite{Hata:1993gf}, Zwiebach and Hata have shown how to relate
slightly different, consistent sets of vertices in CSFT via an infinitesimal
field redefinition. Our strategy is now to apply their method to our
problem in OSFT and integrate the result to the finite case. This
will not only provide us the field redefinition we are looking for,
but also give some insight into the rather abstract formalism of CSFT.
First, it is useful to collect some basic information about the structure
of CSFT.

As already explained in the introduction, the vertices are given by
integrating \emph{basic differential forms} $\Omega_{\Psi_{1}\Psi_{2}...\Psi_{n}}^{g,n}$
defined by 
\begin{equation}
\Omega_{\Psi_{1}\Psi_{2}...\Psi_{n}}^{g,n}\left(\hat{V}_{1},\hat{V}_{2},...,\hat{V}_{6g-6+2n}\right)=\left(2\pi i\right)^{-\left(3g-3+n\right)}\Bra{\Sigma}\mathbf{b\left(v_{1}\right)}...\mathbf{b\left(v_{6g-6+2n}\right)}\Ket{\Psi_{1}}...\Ket{\Psi_{n}}.
\end{equation}
They are living in the tangent space of the fibre bundle $\mathcal{\hat{P}}_{g,n}$
over the moduli space $\mathcal{M}_{g,n}$, with the fiber being the
space of local coordinates around the punctures modulo phase rotations.
The dimension of this bundle is infinite, but the degree of $\Omega_{\Psi_{1}\Psi_{2}...\Psi_{n}}^{g,n}$
is just the real dimension of the base space $\mathcal{M}_{g,n}$.
It takes as arguments tangent vectors $\hat{V}_{i}\in T\mathcal{\hat{P}}_{g,n}$,
which represent deformations of the world sheet Riemann surface $\Sigma$,
either by changing the moduli or the local coordinates. The $\mathbf{v_{i}}$
are Schiffer vectors on $\Sigma$, supported around the punctures,
which generate those deformations. This means that the local coordinate
around the $n$th puncture transforms as 
\begin{equation}
z^{\left(n\right)}\rightarrow z^{\left(n\right)}+\epsilon v^{\left(n\right)}\left(z^{\left(n\right)}\right)\label{eq:Schiffer vector}
\end{equation}
for some small $\epsilon$. The $b$-ghost insertions are then defined
as 
\begin{equation}
\mathbf{b\left(v\right)}=\sum_{i=1}^{n}\left(\oint\frac{dz_{i}}{2\pi i}b\left(z_{i}\right)v^{\left(i\right)}\left(z_{i}\right)+\oint\frac{d\bar{z}_{i}}{2\pi i}\bar{b}\left(\bar{z}_{i}\right)\bar{v}^{\left(i\right)}\left(\bar{z}_{i}\right)\right).
\end{equation}
We will be only interested in the classical action without the loop
vertices, so the genus $g$ shall be zero from now on. The basic forms
can now be integrated over sections of $\mathcal{\hat{P}}_{0,n}$
defining the vertices $\mathcal{V}{}_{0,n}$.

The quantization procedure makes use of the Batalin-Vilkovisky formalism;
although we are not interested in quantum effects, the BV-antibracket
is used in constructing the symmetry generator. It is defined as 
\begin{equation}
\left\{ A,B\right\} =\frac{\partial_{r}A}{\partial\Psi^{i}}\frac{\partial_{l}B}{\partial\Psi_{i}^{*}}-\frac{\partial_{r}A}{\partial\Psi_{i}^{*}}\frac{\partial_{l}B}{\partial\Psi^{i}},
\end{equation}
where the $\Psi_{i}^{*}$ are antifields of opposite parity associated
to each basis element of $\mathcal{H}$. The BV-master action takes
the same form as (\ref{eq:CSFT action}) with the only difference
that the $\Psi$ are not restricted in ghost number and run over fields
as well as antifields.

\subsection{Constructing the field redefinition}

We are looking for a non-linear field redefinition of the form 
\begin{equation}
\Psi'=F\left(\Psi\right)=\stackrel[n=1]{\infty}{\sum}F_{n}\left(\Psi^{\otimes n}\right)=\pi_{1}\mathbf{F}\frac{1}{1-\Psi}\label{eq:finite field redef}
\end{equation}
that relates the Witten action to the stubbed $A_{\infty}$-action
in the form (\ref{eq:deformed action}). To be consistent with the
results of Zwiebach and Hata \cite{Hata:1993gf} we demand 
\begin{equation}
S\left(\Psi'\right)=S'\left(\Psi\right).
\end{equation}
Since the kinetic term is identical, we immediately find 
\begin{equation}
F_{1}\left(\Psi\right)=\Psi.
\end{equation}

In \cite{Hata:1993gf} it is shown that under an infinitesimal field
redefinition of the form 
\begin{equation}
\Psi\rightarrow\Psi+t\left\{ \Psi,e\right\} +\mathcal{O}\left(t^{2}\right),\label{eq:infinitesimal redef}
\end{equation}
the classical action transforms as 
\begin{equation}
S\left(\Psi\right)\rightarrow S\left(\Psi\right)+t\left\{ S,e\right\} +\mathcal{O}\left(t^{2}\right),
\end{equation}
where $\left\{ \right\} $ denotes the BV-antibracket. In the paper
it is now argued that for any small change of vertices, the change
of the action indeed takes this form and the generator $e$ is constructed
explicitly: 
\begin{equation}
e\left(u_{0}\right)=-\sum_{n}\kappa^{n-2}\frac{1}{n!}\underset{\mathcal{V}_{0,n}\left(u_{0}\right)}{\int}\Omega_{\mathbf{b\left(u\right)}\Psi^{\otimes n}}^{\left(0\right)0,n}.
\end{equation}
Here we assume that there exists some family of consistent vertex
sets $\mathcal{V}_{0,n}\left(u\right)$ parametrized by some real
number $u$ and everything is evaluated at the point $u_{0}$. The
vector $\mathbf{u}$ is some Schiffer vector which generates a deformation
of the $\mathcal{V}_{0,n}\left(u_{0}\right)$ in the direction of
$u$, i. e. it generates diffeomorphisms which push $\mathcal{V}_{0,n}\left(u_{0}\right)$
into $\mathcal{V}_{0,n}\left(u_{0}+\delta u\right)$. For the case
of varying the stub length, this Schiffer vector takes a particular
simple form: First, lets notice that the stub length $\lambda$ for
closed strings is defined as the geodesic distance from the location
$\mid z\mid=1$ of the local coordinate to the begin of the semiinfinite
cylinder associated with the puncture. This implies that $\lambda$
can be changed by just rescaling the coordinate: Sending $z$ to $z'=z+\epsilon z$,
the location of $\mid z'\mid=1$ corresponds to $\mid z\mid=1-\epsilon$,
such that $\lambda$ is increased by $\epsilon.$ By comparing with
Eq. (\ref{eq:Schiffer vector}) we read off 
\begin{equation}
u^{\left(i\right)}=z^{\left(i\right)}.
\end{equation}
The $b$-ghost insertion is then given by 
\begin{equation}
\mathbf{b\left(u\right)}=\sum_{i=1}^{n}\left(\oint\frac{dz_{i}}{2\pi i}z_{i}b\left(z_{i}\right)+\oint\frac{d\bar{z}_{i}}{2\pi i}\bar{z}_{i}\bar{b}\left(\bar{z}_{i}\right)\right)=\sum_{i=1}^{n}b_{0}^{\left(i\right)}+\overline{b}_{0}^{\left(i\right)}=\sum_{i=1}^{n}b_{0}^{+\left(i\right)}.
\end{equation}
We want to use the above expression for $e$ in the context of changing
the stub length for open strings, hence a few modifications and simplifications
are necessary: First, the combinatorial factor $n!$ originates from
total symmetrization of the vertices and is not necessary for open
strings. Second, the insertions of $b_{0}^{+}$ should get replaced
simply by $b_{0}$ since there is no antiholomorphic sector. Moreover,
the string coupling $\kappa$ will be set to one. Now the generator
simplifies to 
\begin{equation}
e\left(\lambda\right)=-\sum_{n}\underset{\mathcal{V}\left(\lambda\right)_{0,n}}{\int}\Omega_{\mathbf{b_{0}}\Psi^{\otimes n}}^{0,n}.\label{eq:e generator}
\end{equation}
If we make the ansatz 
\begin{equation}
\Psi'=\Psi+\delta\lambda\stackrel[n=2]{\infty}{\sum}f_{n}\left(\Psi^{\otimes n}\right)
\end{equation}
as the infinitesimal version of (\ref{eq:finite field redef}), then
the $f_{n}$ are determined as 
\begin{equation}
f_{n}\left(\Psi^{\otimes n}\right)=\left\{ \Psi,e^{\left(n\right)}\right\} ,
\end{equation}
where $\delta\lambda$ plays the role of $t$ in (\ref{eq:infinitesimal redef}).

To find $f_{2}\left(\Psi,\Psi\right)$ we need to consider $e\left(\lambda\right)$
for $n=3$: The vertex $\mathcal{V}\left(\lambda\right)_{0,3}$ is
zero dimensional, so there is no integral and the surface state $\Bra{\Sigma}$
is just the Witten vertex with stubs of length $\lambda$, 
\begin{equation}
\Bra{V_{3}\left(\lambda\right)}=\omega\left(\cdot,M_{2}\left(\cdot,\cdot\right)\right)=\omega\left(e^{-\lambda L_{0}}\cdot,e^{-\lambda L_{0}}\cdot\ast e^{-\lambda L_{0}}\cdot\right).
\end{equation}
Inserting into (\ref{eq:e generator}) yields 
\begin{equation}
e^{\left(3\right)}\left(\lambda,\Psi\right)=-\left(\omega\left(b_{0}\Psi,M_{2}\left(\Psi,\Psi\right)\right)+\omega\left(\Psi,M_{2}\left(b_{0}\Psi,\Psi\right)\right)+\omega\left(\Psi,M_{2}\left(\Psi,b_{0}\Psi\right)\right)\right).
\end{equation}
The BV-bracket with $\Psi$ can be straightforwardly evaluated; after
carefully checking the signs the result is

\begin{equation}
f_{2}\left(\Psi,\Psi\right)=-b_{0}M_{2}\left(\Psi,\Psi\right)+M_{2}\left(b_{0}\Psi,\Psi\right)+M_{2}\left(\Psi,b_{0}\Psi\right).
\end{equation}
We expect now the relation 
\begin{equation}
S'\left(\Psi+\delta\lambda f_{2}\left(\Psi,\Psi\right),\lambda\right)=S'\left(\Psi,\lambda+\delta\lambda\right)
\end{equation}
to hold up to order 3 in $\Psi$; by directly inserting we can compute
explicitly 
\begin{align}
S'\left(\Psi+\delta\lambda f_{2}\left(\Psi,\Psi\right),\lambda\right) & =\frac{1}{2}\omega\left(\Psi,Q\Psi\right)+\frac{1}{3}\omega\left(\Psi,M_{2}\left(\Psi,\Psi\right)\right)+\delta\lambda\,\omega\left(\Psi,Qf_{2}\left(\Psi,\Psi\right)\right)+\mathcal{O}\left(\Psi^{\otimes4}\right).
\end{align}
The last and most interesting term yields 
\begin{align}
 & \left(-\omega\left(\Psi,Qb_{0}M_{2}\left(\Psi,\Psi\right)\right)+\omega\left(\Psi,QM_{2}\left(b_{0}\Psi,\Psi\right)\right)+\omega\left(\Psi,QM_{2}\left(\Psi,b_{0}\Psi\right)\right)\right)\delta\lambda\nonumber \\
= & \left(-\omega\left(\Psi,L_{0}M_{2}\left(\Psi,\Psi\right)\right)+\omega\left(\Psi,b_{0}QM_{2}\left(\Psi,\Psi\right)\right)-\omega\left(Q\Psi,M_{2}\left(b_{0}\Psi,\Psi\right)\right)-\omega\left(Q\Psi,M_{2}\left(\Psi,b_{0}\Psi\right)\right)\right)\delta\lambda\nonumber \\
= & -\omega\left(\Psi,L_{0}M_{2}\left(\Psi,\Psi\right)\right)\delta\lambda,
\end{align}
where the last three terms in the second line cancel after applying
the Leibniz rule and cyclicity. On the other hand, 
\begin{align}
S'\left(\Psi,\lambda+\delta\lambda\right)= & \,S'\left(\Psi,\lambda\right)+\delta\lambda\frac{d}{d\lambda}S'\left(\Psi,\lambda\right)=\frac{1}{2}\omega\left(\Psi,Q\Psi\right)+\frac{1}{3}\omega\left(\Psi,M_{2}\left(\Psi,\Psi\right)\right)\nonumber \\
 & -\frac{1}{3}\delta\lambda\,\omega\left(L_{0}\Psi,M_{2}\left(\Psi,\Psi\right)\right)-\frac{1}{3}\delta\lambda\,\omega\left(\Psi,M_{2}\left(L_{0}\Psi,\Psi\right)\right)-\frac{1}{3}\delta\lambda\,\omega\left(\Psi,M_{2}\left(\Psi,L_{0}\Psi\right)\right)+\mathcal{O}\left(\Psi^{\otimes4}\right)\nonumber \\
= & \,\frac{1}{2}\omega\left(\Psi,Q\Psi\right)+\frac{1}{3}\,\omega\left(\Psi,M_{2}\left(\Psi,\Psi\right)\right)-\delta\lambda\,\omega\left(L_{0}\Psi,M_{2}\left(\Psi,\Psi\right)\right)+\mathcal{O}\left(\Psi^{\otimes4}\right),
\end{align}
which is the same expression up to order $\Psi^{\otimes3}$ (Again,
cyclicity was used in the last line.).

The explicit form of $f_{2}\left(\Psi,\Psi\right)$ suggests the following
general structure: We can guess the ansatz 
\begin{equation}
f_{n}\left(\Psi^{\otimes n}\right)=-b_{0}M_{n}\left(\Psi^{\otimes n}\right)+M_{n}\left(\mathbf{b_{0}}\left(\Psi^{\otimes n}\right)\right),\label{eq:f_n in general}
\end{equation}
where $\mathbf{b_{0}}$ again denotes the coderivation associated
to $b_{0}$. At first sight, Eq.(\ref{eq:f_n in general}) looks a
bit strange now from the coalgebra perspective because it is a commutator
of two odd objects, so one would more naturally expect an anticommutator.
However, the first term in (\ref{eq:f_n in general}) stems from the
application of $b_{0}$ on the first argument of the symplectic form
$\omega$, hence the sign contains implicit information about $\omega$.
From the discussion about $\mathbf{P}$ from the HPL we could anticipate
that $\omega$ has to enter the calculation at some point. The more
natural looking expression $\pi_{1}\left[\mathbf{b_{0}},\mathbf{M_{n}}\right]$
would have been independent of $\omega$.

One can prove now that the ansatz (\ref{eq:f_n in general}) is indeed
correct by directly inserting into the action.

\subsection{Proof of the ansatz for the infinitesimal field redefinition}

If we focus solely on terms of order $n+1$ in $\Psi$ we get 
\begin{equation}
S'^{\left(n+1\right)}\left(\Psi'\right)=\omega\left(\Psi,Qf_{n}\left(\Psi^{\otimes n}\right)\right)\delta\lambda\,+\,\sum_{k=2}^{n-1}\omega\left(\Psi,M_{k}\left(f_{n+1-k}\left(\Psi^{\otimes n+1-k}\right),\Psi^{\otimes k-1}\right)\right)\delta\lambda\,+\,\frac{1}{n+1}\omega\left(\Psi,M_{n}\left(\Psi^{\otimes n}\right)\right).
\end{equation}
The last term is the contribution from the original $S\left(\Psi\right)$,
so the first two terms denoted by $\delta S^{\left(n+1\right)}$ should
yield the infinitesimal variation 
\begin{equation}
\delta S'^{\left(n+1\right)}\overset{?}{=}\delta\lambda\frac{d}{d\lambda}\frac{1}{n+1}\omega\left(\Psi,M_{n}\left(\Psi^{\otimes n}\right)\right).\label{eq:lambda derivative}
\end{equation}
Inserting the ansatz and performing some straight forward manipulations
gives 
\begin{align}
\frac{\delta S'^{\left(n+1\right)}}{\delta\lambda}= & \,\omega\left(\Psi,Q\left(-b_{0}M_{n}+M_{n}\mathbf{b_{0}}\right)\left(\Psi^{\otimes n}\right)\right)+\sum_{k=2}^{n-1}\omega\left(\Psi,M_{k}\left(\left(-b_{0}M_{n+1-k}+M_{n+1-k}\mathbf{b_{0}}\right)\left(\Psi^{\otimes n+1-k}\right),\Psi^{\otimes k-1}\right)\right)\nonumber \\
= & \,-\omega\left(\Psi,L_{0}M_{n}\left(\Psi^{\otimes n}\right)\right)+\omega\left(\Psi,b_{0}\left[\mathbf{Q,M_{n}}\right]\left(\Psi^{\otimes n}\right)\right)-\omega\left(\Psi,b_{0}M_{n}\mathbf{Q}\left(\Psi^{\otimes n}\right)\right)+\omega\left(\Psi,QM_{n}\mathbf{b_{0}}\left(\Psi^{\otimes n}\right)\right)\nonumber \\
 & \,+\sum_{k=2}^{n-1}\left(-\omega\left(\Psi,M_{k}\left(b_{0}M_{n+1-k}\left(\Psi^{\otimes n+1-k}\right),\Psi^{\otimes k-1}\right)\right)+\omega\left(\Psi,M_{k}\left(M_{n+1-k}\mathbf{b_{0}}\left(\Psi^{\otimes n+1-k}\right),\Psi^{\otimes k-1}\right)\right)\right).\label{eq:delta S}
\end{align}
The last two terms of the second line actually cancel each other:
\begin{align}
 & -\omega\left(\Psi,b_{0}M_{n}\mathbf{Q}\left(\Psi^{\otimes n}\right)\right)+\omega\left(\Psi,QM_{n}\mathbf{b_{0}}\left(\Psi^{\otimes n}\right)\right)\nonumber \\
= & -\omega\left(b_{0}\Psi,M_{n}\mathbf{Q}\left(\Psi^{\otimes n}\right)\right)-\omega\left(Q\Psi,M_{n}\mathbf{b_{0}}\left(\Psi^{\otimes n}\right)\right)\nonumber \\
= & -\omega\left(b_{0}\Psi,M_{n}\left(Q\Psi,\Psi,...,\Psi\right)\right)-\omega\left(b_{0}\Psi,M_{n}\left(\Psi,Q\Psi,...,\Psi\right)\right)-...-\omega\left(b_{0}\Psi,M_{n}\left(\Psi,\Psi,...,Q\Psi\right)\right)\nonumber \\
 & -\omega\left(Q\Psi,M_{n}\left(b_{0}\Psi,\Psi,...,\Psi\right)\right)-\omega\left(Q\Psi,M_{n}\left(\Psi,b_{0}\Psi,...,\Psi\right)\right)-...-\omega\left(Q\Psi,M_{n}\left(\Psi,\Psi,...,b_{0}\Psi\right)\right).
\end{align}
Because of cyclicity of the $\left(n+1\right)$-vertex the last two
lines contain the same terms, just differing by a sign which comes
from commuting $Q$ with $b_{0}$. Therefore they add to zero. The
second term in the second line of (\ref{eq:delta S}) can be further
manipulated using the $A_{\infty}$-relations:

\begin{align}
\frac{\delta S'^{\left(n+1\right)}}{\delta\lambda}= & -\omega\left(\Psi,L_{0}M_{n}\left(\Psi^{\otimes n}\right)\right)-\sum_{k=2}^{n-1}\omega\left(\Psi,M_{k}\left(b_{0}M_{n+1-k}\left(\Psi^{\otimes n+1-k}\right),\Psi^{\otimes k-1}\right)\right)\nonumber \\
 & +\sum_{k=2}^{n-1}\left(\omega\left(\Psi,M_{k}\left(M_{n+1-k}\left(\mathbf{b_{0}}\left(\Psi^{\otimes n+1-k}\right)\right),\Psi^{\otimes k-1}\right)\right)-\omega\left(\Psi,b_{0}\mathbf{M_{k}M_{n+1-k}}\left(\Psi^{\otimes n}\right)\right)\right).\label{eq:complicated}
\end{align}
Now again, the terms in the last line cancel after using cyclicity:
\begin{align}
 & \omega\left(\Psi,b_{0}\mathbf{M_{k}M_{n+1-k}}\left(\Psi^{\otimes n}\right)\right)=\omega\left(b_{0}\Psi,M_{k}\mathbf{M_{n+1-k}}\left(\Psi^{\otimes n}\right)\right)\nonumber \\
=\, & \omega\left(M_{k}\left(b_{0}\Psi,M_{n+1-k}\left(\Psi^{\otimes n+1-k}\right),\Psi^{\otimes k-2}\right),\Psi\right)+\omega\left(M_{k}\left(b_{0}\Psi,\Psi,M_{n+1-k}\left(\Psi^{\otimes n+1-k}\right),\Psi^{\otimes k-3}\right),\Psi\right)\nonumber \\
 & +...+\omega\left(M_{k}\left(b_{0}\Psi,\Psi^{\otimes k-2},M_{n+1-k}\left(\Psi^{\otimes n+1-k}\right)\right),\Psi\right)+\omega\left(M_{k}\left(b_{0}\Psi,\Psi^{\otimes k-1}\right),M_{n+1-k}\left(\Psi^{\otimes n+1-k}\right)\right).
\end{align}
All terms except for the last one can be further manipulated using
the antisymmetry of $\omega$ and cyclicity of $M_{k}$. For example,
\begin{align}
 & \omega\left(M_{k}\left(b_{0}\Psi,M_{n+1-k}\left(\Psi^{\otimes n+1-k}\right),\Psi^{\otimes k-2}\right),\Psi\right)=-\omega\left(\Psi,M_{k}\left(b_{0}\Psi,M_{n+1-k}\left(\Psi^{\otimes n+1-k}\right),\Psi^{\otimes k-2}\right)\right)\nonumber \\
=\, & \omega\left(M_{k}\left(\Psi,b_{0}\Psi,M_{n+1-k}\left(\Psi^{\otimes n+1-k}\right),\Psi^{\otimes k-3}\right),\Psi\right)=...=\omega\left(M_{k}\left(\Psi^{\otimes k-1},b_{0}\Psi\right),M_{n+1-k}\left(\Psi^{\otimes n+1-k}\right)\right),
\end{align}
so in all of the terms the $M_{n+1-k}$ can be moved to the outermost
right. After all, the terms can be summed up as 
\begin{align}
 & \omega\left(M_{k}\left(\Psi^{\otimes k-1},b_{0}\Psi\right),M_{n+1-k}\left(\Psi^{\otimes n+1-k}\right)\right)+\omega\left(M_{k}\left(\Psi^{\otimes k-2},b_{0}\Psi,\Psi\right),M_{n+1-k}\left(\Psi^{\otimes n+1-k}\right)\right)\nonumber \\
 & +...+\omega\left(M_{k}\left(b_{0}\Psi,\Psi^{\otimes k-1}\right),M_{n+1-k}\left(\Psi^{\otimes n+1-k}\right)\right)\nonumber \\
=\, & \omega\left(M_{k}\left(\mathbf{b_{0}}\left(\Psi^{\otimes k}\right)\right),M_{n+1-k}\left(\Psi^{\otimes n+1-k}\right)\right)\nonumber \\
=\, & \omega\left(\Psi,M_{n+1-k}\left(M_{k}\left(\mathbf{b_{0}}\left(\Psi^{\otimes k}\right)\right),\Psi^{\otimes n-k}\right)\right),
\end{align}
which is after the summation over $k$ identical to the first term
in the second line of (\ref{eq:complicated}), just with opposite
sign. So we arrive at the expression 
\begin{equation}
\frac{\delta S'^{\left(n+1\right)}}{\delta\lambda}=-\omega\left(\Psi,L_{0}M_{n}\left(\Psi^{\otimes n}\right)\right)-\sum_{k=2}^{n-1}\omega\left(\Psi,M_{k}\left(b_{0}M_{n+1-k}\left(\Psi^{\otimes n+1-k}\right),\Psi^{\otimes k-1}\right)\right),\label{eq:delta S final}
\end{equation}
which should now be compared to the result of formula (\ref{eq:lambda derivative}).

The derivative with respect to $\lambda$ can act on the stubs as
well as on the homotopy $h.$ The action on $e^{-\lambda L_{0}}$
inserts a factor of $-L_{0}$ on every input string field of the $\left(n+1\right)$-vertex.
Since the vertices are cyclically symmetric, we get $n+1$ identical
terms, which cancels the prefactor $\frac{1}{n+1}.$ The result is
\begin{equation}
\frac{d}{d\lambda}\frac{1}{n+1}\omega\left(\Psi,M_{n}\left(\Psi^{\otimes n}\right)\right)\supset-\omega\left(L_{0}\Psi,M_{n}\left(\Psi^{\otimes n}\right)\right),
\end{equation}
which is equal to the first term of (\ref{eq:delta S final}). To
compute the action on $h$, the tree representation turns out to be
useful again: First of all, 
\begin{equation}
\frac{d}{d\lambda}h=-2b_{0}e^{-2\lambda L_{0}},
\end{equation}
hence we get a sum of all possible tree diagrams with one propagator
replaced by $-2b_{0}e^{-2\lambda L_{0}}$. We can cut through the
diagram along this replaced propagator and think of the factor $e^{-2\lambda L_{0}}$
as arising from two $e^{-\lambda L_{0}}$-stubs from the leaf and
the root of the two subtrees. Both subtrees are now part of a higher
product $M_{k}$ for some $k$, $2\leq k\leq n-1$. So the whole expression
can be written as a combination of two higher products $M_{k}$, $M_{n+1-k}$
with a factor $-2b_{0}$ inserted: 
\begin{align}
\frac{d}{d\lambda}\frac{1}{n+1}\omega\left(\Psi,M_{n}\left(\Psi^{\otimes n}\right)\right)\supset-\frac{1}{n+1}\sum_{k=2}^{n-1} & \omega\left(\Psi,M_{k}\left(2b_{0}M_{n+1-k}\left(\Psi^{\otimes n+1-k}\right),\Psi^{\otimes k-1}\right)\right)+\nonumber \\
 & \omega\left(\Psi,M_{k}\left(\Psi,2b_{0}M_{n+1-k}\left(\Psi^{\otimes n+1-k}\right),\Psi^{\otimes k-2}\right)\right)+...+\nonumber \\
 & \omega\left(\Psi,M_{k}\left(\Psi^{\otimes k-1},2b_{0}M_{n+1-k}\left(\Psi^{\otimes n+1-k}\right)\right)\right).
\end{align}
Because of cyclicity of the $k+1$-vertex, the different lines contain
the same terms so we have 
\begin{equation}
\frac{d}{d\lambda}\frac{1}{n+1}\omega\left(\Psi,M_{n}\left(\Psi^{\otimes n}\right)\right)\supset-\frac{1}{n+1}\sum_{k=2}^{n-1}2k\cdot\omega\left(\Psi,M_{k}\left(b_{0}M_{n+1-k}\left(\Psi^{\otimes n+1-k}\right),\Psi^{\otimes k-1}\right)\right).\label{eq:term with sum}
\end{equation}

The last bracket can be further manipulated: 
\begin{align}
\omega\left(\Psi,M_{k}\left(b_{0}M_{n+1-k}\left(\Psi^{\otimes n+1-k}\right),\Psi^{\otimes k-1}\right)\right) & =-\omega\left(M_{k}\left(b_{0}M_{n+1-k}\left(\Psi^{\otimes n+1-k}\right),\Psi^{\otimes k-1}\right),\Psi\right)\nonumber \\
=\omega\left(b_{0}M_{n+1-k}\left(\Psi^{\otimes n+1-k}\right),M_{k}\left(\Psi^{\otimes k}\right)\right) & =-\omega\left(M_{n+1-k}\left(\Psi^{\otimes n+1-k}\right),b_{0}M_{k}\left(\Psi^{\otimes k}\right)\right)\nonumber \\
=\omega\left(\Psi,M_{n+1-k}\left(\Psi^{\otimes n-k},b_{0}M_{k}\left(\Psi^{\otimes k}\right)\right)\right) & =\omega\left(\Psi,M_{n+1-k}\left(b_{0}M_{k}\left(\Psi^{\otimes k}\right),\Psi^{\otimes n-k}\right)\right).
\end{align}
In the last step cyclicity of the $\left(n+1-k\right)$-vertex was
used again. We see that in the sum of (\ref{eq:term with sum}) the
$k$th term and the $\left(n+1-k\right)$th term are identical so
the sum can be rewritten as 
\begin{align}
 & -\frac{1}{n+1}\sum_{k=2}^{n-1}\left(n+1\right)\cdot\omega\left(\Psi,M_{k}\left(b_{0}M_{n+1-k}\left(\Psi^{\otimes n+1-k}\right),\Psi^{\otimes k-1}\right)\right)\nonumber \\
= & \sum_{k=2}^{n-1}\omega\left(\Psi,M_{k}\left(b_{0}M_{n+1-k}\left(\Psi^{\otimes n+1-k}\right),\Psi^{\otimes k-1}\right)\right),
\end{align}
which is precisely the second term in (\ref{eq:delta S final}).

This completes the proof that the ansatz 
\begin{equation}
f_{n}\left(\Psi^{\otimes n}\right)=-b_{0}M_{n}\left(\Psi^{\otimes n}\right)+M_{n}\left(\mathbf{b_{0}}\left(\Psi^{\otimes n}\right)\right)
\end{equation}
indeed yields the correct infinitesimal field redefinition.

\subsection{Finite field redefinition}

So far we have only been concerned with infinitesimal variations of
$\lambda$, now we want to generalize the results to finite changes.
We know 
\begin{equation}
\Psi_{\lambda+\delta\lambda}=\Psi_{\lambda}+\delta\lambda f_{2}^{\lambda}\left(\Psi_{\lambda},\Psi_{\lambda}\right)+\delta\lambda f_{3}^{\lambda}\left(\Psi_{\lambda},\Psi_{\lambda},\Psi_{\lambda}\right)+...=\Psi_{\lambda}+\delta\lambda\frac{d}{d\lambda}\Psi_{\lambda},
\end{equation}
where we have written the superscript $\lambda$ to indicate that
the $f_{n}$ also depend on $\lambda$ explicitly. This equation can
be integrated to 
\begin{equation}
\Psi_{\lambda}=\Psi_{0}+\int_{0}^{\lambda}dtf_{2}^{t}\left(\Psi_{t},\Psi_{t}\right)+\int_{0}^{\lambda}dtf_{3}^{t}\left(\Psi_{t},\Psi_{t},\Psi_{t}\right)+...\,\,\,\,\,\,\,\,\,.
\end{equation}
Inserting $\Psi_{\lambda}$ back we get a perturbative expansion in
the original solution $\Psi_{0}:$ 
\begin{align}
\Psi_{\lambda}= & \Psi_{0}+\int_{0}^{\lambda}dt\,\,f_{2}^{t}\left(\Psi_{0},\Psi_{0}\right)+\int_{0}^{\lambda}dt\,\,f_{3}^{t}\left(\Psi_{0},\Psi_{0},\Psi_{0}\right)+\nonumber \\
 & \int_{0}^{\lambda}dt\,\,f_{2}^{t}\left(\int_{0}^{t}ds\,\,f_{2}^{s}\left(\Psi_{0},\Psi_{0}\right),\Psi_{0}\right)+\int_{0}^{\lambda}dt\,\,f_{2}^{t}\left(\Psi_{0},\int_{0}^{t}ds\,\,f_{2}^{s}\left(\Psi_{0},\Psi_{0}\right)\right)+\mathcal{O}\left(\Psi_{0}^{\otimes4}\right).
\end{align}
This formula provides an algorithm to find the associated $A_{\infty}$-solution
to each known solution of the Witten OSFT.

\section{Physical interpretation}

To summarize, we found two distinct field redefinitions 
\begin{equation}
\tilde{\Psi}=\stackrel[n=1]{\infty}{\sum}P_{n}\left(\Psi^{\otimes n}\right),\,\,\,\,\,\,\,\,\,\,\Psi'=\stackrel[n=1]{\infty}{\sum}F_{n}\left(\Psi^{\otimes n}\right),
\end{equation}
which generate two different actions $\tilde{S}$ and $S'$ via 
\begin{equation}
\tilde{S}\left(\tilde{\Psi}\right)=S\left(\Psi\right),\,\,\,\,\,\,\,\,\,\,S'\left(\Psi\right)=S\left(\Psi'\right).
\end{equation}
However, both actions share the same equations of motion, namely the
Maurer-Cartan equation of the stubbed $A_{\infty}$-algebra 
\begin{equation}
\pi_{1}\mathbf{M}\frac{1}{1-\Psi}=0.
\end{equation}
It remains to examine what the physical meaning of those two actions
is, most importantly, if they yield the same on-shell value for a
given solution $\Psi^{\ast}$.

In \cite{Mukherji:1991tb,Scheinpflug:2023osi}, a special class of
solutions containing a nearly marginal vertex operator is introduced
which serves as a useful playground to analyze this question.

\subsection{Nearly marginal solutions}

Consider a matter conformal primary field $V$ with weight $h$ smaller
but very close to one. The string field 
\begin{equation}
\Psi_{1}=\mu\cdot cV\left(0\right)\Ket{0},
\end{equation}
with some real coupling constant $\mu$ obeys the Siegel gauge condition
\begin{equation}
b_{0}\Psi_{1}=0
\end{equation}
and will serve as a starting point to find the full solution $\Psi=\sum\Psi_{n}$
as a perturbative series in the expansion parameter $y=1-h$. One
can solve for the string coupling $\mu$ using the Witten equations
of motion, 
\begin{equation}
Q\Psi+\Psi\ast\Psi=0,
\end{equation}
to obtain \cite{Scheinpflug:2023osi} 
\begin{equation}
\mu=\frac{y}{C_{VVV}}+\mathcal{O}\left(y^{3}\right),
\end{equation}
where $C_{VVV}$ denotes the three-point function constant of $V$.
We can deduce that $\Psi_{1}=\mathcal{O}\left(y\right)$ and from
the perturbative algorithm for the full solution one can also show
that in general $\Psi_{n}=\mathcal{O}\left(y^{n}\right)$.

The on-shell action in Witten theory can be written compactly as 
\begin{equation}
S\left(\Psi\right)=-\frac{1}{6}\Braket{\Psi,Q\Psi}.\label{on-shell Witten action}
\end{equation}
From 
\begin{equation}
Q\Psi_{1}=\mu y\cdot c\partial cV\left(0\right)\Ket{0}
\end{equation}
we see that the action will be of leading order $y^{3}$ and given
by 
\begin{equation}
S\left(\Psi\right)=-\frac{1}{6}\frac{y^{3}}{C_{VVV}^{2}}\Braket{cV,c\partial cV}=\frac{1}{6}\frac{y^{3}}{C_{VVV}^{2}}+\mathcal{O}\left(y^{4}\right),
\end{equation}
if we assume that $V$ is conveniently normalized, $\Braket{V\left(z_{1}\right),V\left(z_{2}\right)}=z_{12}^{-2h}$.

\subsection{The $A_{\infty}$-action }

The first important observation is that the cohomomorphism $\mathbf{P}$
simplifies significantly for string fields in Siegel gauge: Since
$h$ is proportional to $b_{0}$, it annihilates $\Psi$ and (\ref{eq:interacting Ps})
collapses to 
\begin{equation}
\tilde{\Psi}=\stackrel[n=1]{\infty}{\sum}P_{n}\left(\Psi^{\otimes n}\right)=p\Psi.
\end{equation}
We already know that $\tilde{S}\left(\tilde{\Psi}\right)$ yields
the original value $S\left(\Psi\right)$, so now we want to check
the expression $S'\left(\tilde{\Psi}\right)$: $\Psi_{1}$ is an $L_{0}$-eigenstate
so we straightforwardly get 
\begin{equation}
p\Psi_{1}=e^{-\lambda L_{0}}\Psi_{1}=e^{\lambda y}\Psi_{1}.
\end{equation}
For cubic order in $y$ we just have to insert this into the kinetic
term (\ref{on-shell Witten action}) and get 
\begin{equation}
S'\left(\tilde{\Psi}\right)\mid_{y^{3}}=-\frac{1}{6}\Braket{p\Psi_{1},Qp\Psi_{1}}=\frac{1}{6}\frac{y^{3}}{C_{VVV}^{2}}e^{2\lambda y}\mid_{y^{3}}=\frac{1}{6}\frac{y^{3}}{C_{VVV}^{2}},
\end{equation}
which agrees with the result above up to terms of $\mathcal{O}\left(y^{4}\right)$.

For a more non-trivial check we can collect the terms of quartic order
in $y$: In the action we have to consider the first three terms 
\begin{equation}
S'\left(\tilde{\Psi}\right)\mid_{y^{4}}=\left(-\frac{1}{2}\Braket{\tilde{\Psi},Q\tilde{\Psi}}-\frac{1}{3}\Braket{\tilde{\Psi},M_{2}\left(\tilde{\Psi},\tilde{\Psi}\right)}-\frac{1}{4}\Braket{\tilde{\Psi},M_{3}\left(\tilde{\Psi},\tilde{\Psi},\tilde{\Psi}\right)}\right)\mid_{y^{4}}.
\end{equation}
The equations of motion however tell us that 
\begin{equation}
\Braket{\tilde{\Psi},Q\tilde{\Psi}}+\Braket{\tilde{\Psi},M_{2}\left(\tilde{\Psi},\tilde{\Psi}\right)}+\Braket{\tilde{\Psi},M_{3}\left(\tilde{\Psi},\tilde{\Psi},\tilde{\Psi}\right)}=\mathcal{O}\left(y^{5}\right),
\end{equation}
so the expression simplifies to 
\begin{equation}
S'\left(\tilde{\Psi}\right)\mid_{y^{4}}=\left(-\frac{1}{4}\Braket{\tilde{\Psi},Q\tilde{\Psi}}-\frac{1}{12}\Braket{\tilde{\Psi},M_{2}\left(\tilde{\Psi},\tilde{\Psi}\right)}\right)\mid_{y^{4}}.
\end{equation}
Plugging in $\tilde{\Psi}=e^{\lambda y}\Psi_{1}$ and isolating $y^{4}$-terms
yields 
\begin{align}
S'\left(\tilde{\Psi}\right)\mid_{y^{4}}= & \left(-\frac{1}{4}\Braket{e^{\lambda y}\Psi_{1},Qe^{\lambda y}\Psi_{1}}-\frac{1}{12}\Braket{e^{\lambda y}\Psi_{1},e^{-\lambda L_{0}}\left(e^{-\lambda L_{0}}e^{\lambda y}\Psi_{1}\ast e^{-\lambda L_{0}}e^{\lambda y}\Psi_{1}\right)}\right)\mid_{y^{4}}\nonumber \\
= & \left(-\frac{1}{4}e^{2\lambda y}\Braket{\Psi_{1},Q\Psi_{1}}-\frac{1}{12}e^{6\lambda y}\Braket{\Psi_{1},\left(\Psi_{1}\ast\Psi_{1}\right)}\right)\mid_{y^{4}}\nonumber \\
= & -\frac{1}{2}\frac{\lambda y^{4}}{C_{VVV}^{2}}\Braket{cV,c\partial cV}-\frac{1}{2}\frac{\lambda y^{4}}{C_{VVV}^{3}}\Braket{cV,\left(cV\ast cV\right)}.\label{eq:S' to order y^4}
\end{align}
The correlation functions can be calculated by using standard CFT
methods, see e.g. \cite{Okawa:2012ica,Scheinpflug:2023osi}; the result
is 
\begin{equation}
\Braket{cV,c\partial cV}=-1,\,\,\,\,\,\,\,\,\,\,\,\,\Braket{cV,\left(cV\ast cV\right)}=C_{VVV}\left(\frac{3\sqrt{3}}{4}\right)^{3y}=C_{VVV}\left(1+3y\cdot\text{ln}\left(\frac{3\sqrt{3}}{4}\right)\right)+\mathcal{O}\left(y^{2}\right).
\end{equation}
We see by inserting into (\ref{eq:S' to order y^4}) that $S'\left(\tilde{\Psi}\right)\mid_{y^{4}}$
indeed vanishes and the value of the on-shell action is the same as
the original $S\left(\Psi\right)$ to order $y^{4}$. In principle,
terms containing $\Psi_{2}\propto y^{2}$ also contribute at this
order. However, since any appearance of $\lambda$ automatically comes
with a factor $y$, there are no terms of order $y^{4}$ containing
$\lambda$ as well as $\Psi_{2}$. All $\Psi_{2}$-contributions are
just the ones already present in $S\left(\Psi\right)$ and were studied
in detail in \cite{Scheinpflug:2023osi}.

We see that to the first two leading orders, the actions $S'\left(\tilde{\Psi}\right)$
and $\tilde{S}\left(\tilde{\Psi}\right)=S\left(\Psi\right)$ give
the same on-shell result. Since $S'\left(\tilde{\Psi}\right)=S\left(\tilde{\Psi}'\right)$,
where 
\begin{equation}
\tilde{\Psi}'=\stackrel[n=1]{\infty}{\sum}F_{n}\left(\tilde{\Psi}^{\otimes n}\right)=\pi_{1}\mathbf{FP}\frac{1}{1-\Psi},
\end{equation}
this suggests that the combination $\mathbf{FP}$ gives rise to a
gauge transformation rather than a physically distinct solution. However,
a full proof of this statement will be left for future publications.

\section{Conclusion and outlook}

We succeeded in providing an explicit consistent description of OSFT
with stubs and interestingly found two possible actions with the same
equations of motion. The field redefinitions used to convert solutions
of Witten OSFT to the new theory are given in explicit form. We hope
that the analysis of solutions to the stubbed equations of motion
can teach us more general properties of solutions to Maurer-Cartan
equations for $A_{\infty}$- or $L_{\infty}$-algebras, in particular
about the solutions of closed string field theory. One way to proceed
would be to transform the whole construction to the sliver frame,
where many analytic solutions of OSFT are formulated.

Another possible future direction is to examine wether the stubbed
theory is ``more well-behaved'' in the sense that some typical singularities
and ambiguities, for example connected to identity-like solutions,
are ameliorated.

\subsubsection*{Acknowledgements}

We thank Ted Erler, Jakub Vošmera, Branislav Jur\v{c}o, Igor Khavkine
and Martin Markl for useful discussions. Our work has been funded
by the Grant Agency of Czech Republic under the grant EXPRO 20-25775X.

\appendix

\subsection*{Appendix}

\subsubsection*{Tensor coalgebras}

The tensor coalgebra $TV$ associated to a (graded) vector space $V$
is defined as the Fock space 
\begin{equation}
V^{\otimes0}+V^{\otimes1}+V^{\otimes2}+...
\end{equation}
together with the comultiplication $\Delta:\,\,TV\rightarrow TV\otimes'TV$
given by 
\begin{equation}
\Delta\left(v_{1}\otimes...\otimes v_{n}\right)=\sum_{k=0}^{n}\left(v_{1}\otimes...\otimes v_{k}\right)\otimes'\left(v_{k+1}\otimes...\otimes v_{n}\right)
\end{equation}
on homogeneous elements and extended by linearity. Here the $v_{i}$
are elements of $V$ and $\otimes'$ denotes the tensor product arising
from a comultiplication, in contrast to the usual $\otimes$. We define
the projection operator $\pi_{n}:\,\,TV\rightarrow TV$ to project
any element on its $n$th tensor power component, 
\begin{equation}
\pi_{n}TV=V^{\otimes n}.
\end{equation}

A linear map $\mathbf{d}:\,\,TV\rightarrow TV$ is called a coderivation
if it satisfies the co-Leibniz rule: 
\begin{equation}
\Delta\mathbf{d}=\left(\mathbf{d}\otimes'\mathbf{1}+\mathbf{1}\otimes'\mathbf{d}\right)\Delta.\label{eq:co leibniz}
\end{equation}
Linear combinations of coderivations are again coderivations as well
as their graded commutator 
\begin{equation}
\left[\mathbf{d}_{1},\mathbf{d}_{2}\right]=\mathbf{d}_{1}\mathbf{d}_{2}-\left(-1\right)^{deg\left(\mathbf{d}_{1}\right)deg\left(\mathbf{d}_{2}\right)}\mathbf{d}_{2}\mathbf{d}_{1}.
\end{equation}
The product $\mathbf{d}_{1}\mathbf{d}_{2}$ is in general not a coderivation.
For any $m$-linear map $d_{m}:\,\,V^{\otimes m}\rightarrow V$ one
can construct an associated coderivation by the formula 
\begin{equation}
\mathbf{d}=\underset{n=m}{\overset{\infty}{\sum}}\underset{k=0}{\overset{n-m}{\sum}}1^{\otimes k}\otimes d_{m}\otimes1^{\otimes n-k-m}.\label{eq:coderivation}
\end{equation}
The co-Leibniz rule guarantees that any coderivation is a sum of terms
of this form for different $m$. The individual $m$-products can
be recovered as 
\begin{equation}
d_{m}=\pi_{1}\mathbf{d}\pi_{m}.
\end{equation}
If an odd coderivation $\mathbf{d}$ obeys 
\begin{equation}
\mathbf{d}^{2}=0
\end{equation}
then its components $d_{m}$ form an $A_{\infty}$-algebra.

A linear map $\mathbf{f}$ is called a cohomomorphism if it fulfills
\begin{equation}
\Delta\mathbf{f}=\left(\mathbf{f}\otimes'\mathbf{f}\right)\Delta.
\end{equation}
Linear combinations and products of cohomomorphisms are again cohomomorphisms.
Given a family of $m$-products $f_{m}$ one can construct a unique
cohomomorphism via 
\begin{equation}
\mathbf{f}=\sum_{j=1}^{\infty}\sum_{k=1}^{\infty}\sum_{m_{1}+...+m_{j}=k}f_{m_{1}}\otimes...\otimes f_{m_{j}}.
\end{equation}
Again, the individual products can be recovered from $\mathbf{f}$
as 
\begin{equation}
f_{m}=\pi_{1}\mathbf{f}\pi_{m}.
\end{equation}

Of special importance are elements of $TV$ of the form 
\begin{equation}
1+v+v\otimes v+v\otimes v\otimes v+...=:\frac{1}{1-v}
\end{equation}
for some $v\in V.$ They fulfill the following useful properties:
\begin{equation}
\pi_{1}\mathbf{f}\frac{1}{1-v}=\sum_{m=1}^{\infty}f_{m}\left(v^{\otimes m}\right),
\end{equation}
\begin{equation}
\mathbf{f}\frac{1}{1-v}=\frac{1}{1-\pi_{1}\mathbf{f}\frac{1}{1-v}}\label{eq:cohomo relation}
\end{equation}
for any cohomomorphism $\mathbf{f}.$

A bilinear map $\Bra{\omega}$: $TV\times TV\rightarrow\mathbb{C}$
is called a symplectic form if it satisfies 
\begin{equation}
\Bra{\omega}v_{1}\otimes v_{2}=:\omega\left(v_{1},v_{2}\right)=-\left(-1\right)^{deg\left(v_{1}\right)deg\left(v_{2}\right)}\omega\left(v_{2},v_{1}\right).
\end{equation}
A multilinear product $m_{k}$ is called cyclic with respect to $\omega$
if it fulfills 
\begin{equation}
\omega\left(v_{1},m_{k}\left(v_{2},...,v_{k+1}\right)\right)=-\left(-1\right)^{deg\left(v_{1}\right)deg\left(m_{k}\right)}\omega\left(m_{k}\left(v_{1},...,v_{k}\right),v_{k+1}\right).
\end{equation}
A coderivation $\mathbf{d}$ is cyclic if all of its components $d_{m}=\pi_{1}\mathbf{d}\pi_{m}$
are cyclic or equivalently 
\begin{equation}
\Bra{\omega}\pi_{2}\mathbf{d}=0.
\end{equation}
Given two symplectic forms $\Bra{\omega}$, $\Bra{\omega'}$, a cohomomorphism
$\mathbf{f}$ is cyclic if 
\begin{equation}
\Bra{\omega'}\pi_{2}\mathbf{f}=\Bra{\omega}\pi_{2}.
\end{equation}


\end{document}